\documentclass[twocolumn,aps,superscriptaddress,showpacs,nofootinbib]{revtex4}
\usepackage{hyperref}
\usepackage{footnote}
\usepackage{amssymb}
\usepackage{amsmath}
\usepackage{graphicx}
\usepackage[normalem]{ulem}
\usepackage{url}
\usepackage{csquotes}
\usepackage{color}

\begin{document}
\title{Statistical approach of nuclear multifragmentation with realistic nuclear equation of state}

\author{S. Mallik}
\email{swagato@vecc.gov.in}
\affiliation{Physics Group, Variable Energy Cyclotron Centre, 1/AF Bidhan Nagar, Kolkata 700064, India}
\affiliation{Homi Bhabha National Institute, Training School Complex, Anushakti Nagar, Mumbai 400085, India}

\begin{abstract}
In this work, Canonical Thermodynamical model for nuclear multifragmentation has been updated with realistic nuclear equation of state. Mass distribution, intermediate mass fragment multiplicity as well as isospin sensitive observables have been investigated with semi-microscopic approach of determining nuclear binding and excitation energies. Production of neutron rich isotopes as well as isoscaling and isobaric yield ratio parameters have been significantly modified due to inclusion of this realistic nuclear equation of state.
\end{abstract}

\maketitle
\section{Introduction}
The study of nuclear multifragmentation is important to understand the reaction
mechanism in heavy-ion collisions at intermediate and high energies \cite{DasGupta_book,Bao-an-li2}. Nuclear multifragmentation reactions are commonly used study nuclear liquid gas phase transition \cite{DasGupta_book,Siemens,Gross_phase_transition,Bondorf1,Dasgupta_Phase_transition,Chomaz,Borderie2,Borderie}, nuclear equation of state \cite{Bao-an-li2,Bao-an-li1} and to extrapolate the thermodynamic properties of astrophysical environment of worm stellar matter from laboratory scenario \cite{Botvina_multi_astro,Francesca_multi_astro}. Statistical approaches are quite successful for studying nuclear multifragmentation reactions at intermediate energies. The disintegration of excited nuclei are commonly studied by implementing of different statistical ensembles. Statistical multifragmentation model proposed by Copenhagen group \cite{Bondorf1}, the microcanonical models of Gross \cite{Gross1} and Randrup and Koonin \cite{Randrup} and Canonical thermodynamical model (CTM) \cite{Das} are widely used. In these statistical models, observables related to nuclear fragment cross-section (and/or multiplicity) are usually determined based on the available phase space calculation. Now, to get the accurate phase space, nuclear binding and excitation of all clusters are required and in most of the existing statistical approaches like Canonical Thermodynamical model, Statistical multifragmentation model etc. binding is determined from Bethe-Weizsacker mass formula which successfully explained the ground state properties at zero temperature and saturation nuclear density \cite{Bethe,Weizsacker}. However nuclear multifragmentation occurs at sub-saturation density and higher excitation energy. To include temperature effect in bulk energy part Fermi gas model is commonly used in statistical models of nuclear multifragmentation and for surface energy part, various additional parametrization in Bethe-Weizsacker mass formula is introduced. The density and/or temperature dependence of the nuclear binding also plays an important role in the study of stellar matter properties. It also has significant influence for determining the nuclear properties extremely neutron rich and neutron-deficient nuclei formed in multifragmentation reactions.\\
\indent
The aim of this work is to implement more realistic binding and excitation in Canonical thermodynamical model (CTM) for nuclear multifragmentation and to study the impact of it on the basic observables like intermediate mass fragment (IMF) multiplicity, mass distribution as well as isospin sensitive observalbes like isotopic distribution, isoscaling, isobaric yield ratio. In order to study the decay of the excited fragments produced in the multifragmentation stage, the evaporation model \cite{Mallik1} is also updated with the same realistic binding and excitation.\\
\indent
The paper is structured as follows. In section II, a brief introduction of the Canonical Thermodynamical model is presented. The results are described in section III, finally summary and conclusions are discussed in section IV.
\section{Model Description}
In CTM, \cite{DasGupta_book,Das} it is assumed that statistical equilibrium is attained at freeze-out stage and population of different channels of disintegration is solely decided by statistical weights in the available phase space. The calculation is done for a fixed mass and atomic number, freeze out volume and temperature. In a canonical model \cite{Das}, the partitioning is done such that all partitions have the correct $A_{0},Z_{0}$ (equivalently $N_{0},Z_{0}$). The canonical partition function is given by
\begin{eqnarray}
Q_{N_{0},Z_{0}} & = & \sum\prod\frac{\omega_{N,Z}^{n_{N,Z}}}{n_{N,Z}!}
\end{eqnarray}
where the sum is over all possible channels of break-up (the number of such channels is enormous) satisfying $N_{0}=\sum N\times n_{N,Z}$ and $Z_{0}=\sum Z\times n_{N,Z}$; $\omega_{N,Z}$ is the partition function of the composite with $N$ neutrons \& $Z$ protons and $n_{NZ}$ is its multiplicity. The partition function $Q_{N_{0},Z_{0}}$ is calculated by applying a recursion relation \cite{Chase}. From Eq. (1), the average number of composites with $N$ neutrons and $Z$ protons can be expressed as,
\begin{eqnarray}
\langle n_{N,Z}\rangle & = & \omega_{N,Z}\frac{Q_{N_{0}-N,Z_{0}-Z}}{Q_{N_{0},Z_{0}}}
\end{eqnarray}
\indent
The partition function of a composite having $N$ neutrons and $Z$ protons is a product of two parts: one is due to the the translational motion and the other is the intrinsic partition function of the composite:
\begin{eqnarray}
\omega_{N,Z}=\frac{V}{h^{3}}(2\pi mT)^{3/2}A^{3/2}\times z_{N,Z}(int)
\end{eqnarray}
where $V$ is the volume available for translational motion, which can be expressed as $V=V_{f}-V_{0}$ , where $V_{0}$ is the normal volume of nucleus with $Z_{0}$ protons and $N_{0}$ neutrons. $z_{N,Z}(int)$ is the internal partition function, the proton and the neutron are fundamental building blocks, thus $z_{1,0}(int)=z_{0,1}(int)=2$, where 2 takes care of the spin degeneracy. For $^2$H, $^3$H, $^3$He, $^4$He, $^5$He and $^6$He, $z_{N,Z}(int)=(2s_{N,Z}+1)\exp[-\beta E_{N,Z}(gr)]$ where $\beta=1/T, E_{N,Z}(gr)$ is the ground-state energy of the composite and $(2s_{N,Z}+1)$ is the experimental spin degeneracy of the ground state.  Excited states for these very low-mass nuclei are not included.\\
\subsection{Conventional approach of of determining internal partition function in CTM}
\indent
In conventional CTM approach as well as in most of the other statistical models of multifragmentation, for determining the internal partition function of nuclei with $Z \geq$3 the liquid-drop formula is used for calculating the binding energy and the contribution for excited states is taken from the Fermi-gas model. In CTM approach, the internal partition function is usually expressed as,
\begin{eqnarray}
z_{N,Z}(int)
&=&\exp\frac{1}{T}\bigg{[}W_0A-a_s(T)A^{2/3}-a^{*}_c\frac{Z^2}{A^{1/3}}\nonumber\\
&&-C_{sym}\frac{(N-Z)^2}{A}+\frac{T^2A}{\epsilon_0}\bigg{]}
\end{eqnarray}
\indent
The expression includes the volume energy [$W_0=15.8$ MeV], the temperature dependent surface energy
[$a_s(T)=a_{s0}\{(T_{c}^2-T^2)/(T_{c}^2+T^2)\}^{5/4}$ with $a_{s0}=18.0$ MeV and $T_{c}=18.0$ MeV], the Coulomb energy [$a^{*}_c=0.31a_{c}$ with $a_{c}=0.72$ MeV and Wigner-Seitz correction factor 0.31 \cite{Bondorf1}] and the symmetry energy ($C_{sym}=23.5$ MeV).  The term $\frac{T^2A}{\epsilon_0}$ ($\epsilon_{0}=16.0$ MeV) represents contribution from excited states since the composites are at a non-zero temperature. In Ref. \cite{Nousad} the volume term of the nuclear binding for extended canonical model is determined from relativistic mean field approach.\\
\subsection{New approach for determining more realistic internal partition function in CTM}
\indent
In this work the internal partition function is determined from semi-microscopic approach where the Helmholtz free energy of a nucleus with $N$ neutrons and $Z$ protons can be decomposed as \cite{Mallik_NuclAstro1},
\begin{equation}
F_{N,Z}=F^{bulk}+F^{surf}+F^{coul} . \label{eq:leptodermous}
\end{equation}
where the bulk part $F^{bulk}$ is originated from bulk nuclear matter at baryonic density $\rho_c=\rho_{c,n}+\rho_{c,z}$ ($\rho_{c,n}$ and $\rho_{c,z}$ are neutron and proton density respectively) and isospin asymmetry $\delta_c= (\rho_{c,n}-\rho_{c,z})/\rho_c=\frac{N-Z}{N+Z}$ occupying a finite spatial volume $V_c=(N+Z)/\rho_c$. The baryonic density with isospin asymmetry $\delta_c$ is approximated \cite{Gulminelli2015}to the corresponding saturation density ($\rho_0$) of symmetric nuclear matter at finite asymmetry according to:
\begin{eqnarray}
\rho_c(\delta_c)=\rho_0\bigg{(}1-\frac{3L_{sym}\delta_c^2}{K_{sat}+K_{sym}\delta_c^2}\bigg{)}.
\end{eqnarray}
Therefore the bulk part of the Helmholtz free energy given by,
\begin{eqnarray}
F^{bulk}=V_c\bigg{[}-\frac{2}{3}\sum_{q=n,z}\xi_{c,q}+\sum_{q=n,z}\rho_{c,q}\eta_q+v(\rho_c,\delta_c)\bigg{]}
\end{eqnarray}
where $\xi_{c,n}$ and $\xi_{c,z}$ are the kinetic energy density of the nucleus due to neutron ($q=n$) and proton ($q=p$) contribution respectively, which can be expressed as $q=n,p$,
\begin{eqnarray}
\xi_{c,q}&=& \frac{3h^2}{2\pi m^*_{c,q}}\bigg{(}\frac{2\pi m^*_{c,q}T}{h^2}\bigg{)}^{5/2}F_{3/2}(\eta_{c,q})
\end{eqnarray}
with
$\eta_{c,q}=F^{-1}_{1/2}\bigg{\{}\bigg{(}\frac{2\pi m^*_{g,q}T}{h^2}\bigg{)}^{3/2}\rho_{c,q}\bigg{\}}$, $F_{1/2}$ and $F_{3/2}$ are the Fermi integrals. The expression of potential energy per particle that can be adapted to different effective interactions and energy functionals  is given by:
\begin{eqnarray}
v(\rho_c,\delta_c)&=&\sum_{k=0}^{N}\frac{1}{k!}(v^{is}_{k}+v^{iv}_{k}\delta_c^2)x^{k}\nonumber\\
&+&(a^{is}+a^{iv}\delta_c^2)x^{N+1}\exp(-b\frac{\rho_c}{\rho_0}) ,
\label{ELFc_potential}
\end{eqnarray}
where $x=\frac{\rho_c-\rho_0}{3\rho_0}$,$a^{is}=-\sum_{k\geq0}^{N}\frac{1}{k!}v^{is}_{k}(-3)^{N+1-k}$ and $a^{iv}=-\sum_{k\geq0}^{N}\frac{1}{k!}v^{iv}_{k}(-3)^{N+1-k}$. $N=4$ and $b=10ln2$ are choosen for this model. This value of $b$ leads to a good reproduction of the Sly5 functional which is used for the numerical applications presented in this paper. The model parameters $v_k^{is(iv)}$ can be linked with a one-to-one correspondence to the usual EoS empirical parameters \cite{Margueron2018}, via:
\begin{eqnarray}
v^{is}_{0}&=&E_{sat}-t_0(1+\kappa_0)\nonumber\\
v^{is}_{1}&=&-t_0(2+5\kappa_0)\nonumber\\
v^{is}_{2}&=&K_{sat}-2t_0(-1+5\kappa_0)\nonumber\\
v^{is}_{3}&=&Q_{sat}-2t_0(4-5\kappa_0)\nonumber\\
v^{is}_{4}&=&Z_{sat}-8t_0(-7+5\kappa_0)
\label{Isoscalar_parameters}
\end{eqnarray}
\begin{eqnarray}
v^{iv}_{0}&=&E_{sym}-\frac{5}{9}t_0[\{1+(\kappa_0+3\kappa_{sym})\}]\nonumber\\
v^{iv}_{1}&=&L_{sym}-\frac{5}{9}t_0[\{2+5(\kappa_0+3\kappa_{sym})\}]\nonumber\\
v^{iv}_{2}&=&K_{sym}-\frac{10}{9}t_0[\{-1+5(\kappa_0+3\kappa_{sym})\}]\nonumber\\
v^{iv}_{3}&=&Q_{sym}-\frac{10}{9}t_0[\{4-5(\kappa_0+3\kappa_{sym})\}]\nonumber\\
v^{iv}_{4}&=&Z_{sym}-\frac{40}{9}t_0[\{-7+5(\kappa_0+3\kappa_{sym})\}] \ ,
\label{Isovector_parameters}
\end{eqnarray}
where  $E_{sat}$, $K_{sat}$, $Q_{sat}$ and $Z_{sat}$ are saturation energy, incompressibility modulus,  isospin symmetric skewness and  kurtosis respectively and $E_{sym}$, $L_{sym}$, $K_{sym}$, $Q_{sym}$ and $Z_{sym}$ are symmetry energy, slope, and associated incompressibility,  skewness and  kurtosis respectively.
Concerning the $\kappa_0$ and $\kappa_{sym}$, they govern the density dependence of the neutron and proton effective mass according to:
\begin{equation}
\frac{m_q}{m^*_{q}(\rho_c,\delta_c)}=1+(\kappa_0 \pm \kappa_{sym}\delta)\frac{\rho_c}{\rho_0},
\end{equation}
with $q=n,p$.
For the applications presented in this paper, all the parameters are taken from the Sly5 functional \cite{Sly5}.\\
The finite size corrections are included by the surface part of the Helmholtz free energy ($F^{surf}$) \cite{papa} for which we adopt the prescription proposed in ref.\cite{LS,Carreau2019,Carreau2019a} on the basis of Thomas-Fermi calculations with extreme isospin ratios:
\begin{eqnarray}
F^{surf}&=&4\pi r^2_c A_N^{2/3}\sigma(y_{c,p},T)
\label{Surface}
\end{eqnarray}
with $r_0=\bigg{\{}\frac{3}{4\pi\rho_0}\bigg{\}}^{1/3}$, $y_{c,p}=Z/(Z+N)$ and
\begin{eqnarray}
\sigma(y_{c,p})=\sigma_0 h\bigg{(}\frac{T}{T_c(y_{c,p})}\bigg{)}\frac{2^{p+1}+b_s}{y_{c,p}^{-p}+b_s+(1-y_{c,p})^{-p}}
\end{eqnarray}
where $\sigma_0$ represents the surface tension of symmetric nuclear matter and $b_s$ and $p$ represent the isospin dependence. For Sly5 functional the parameters were optimized in \cite{Carreau2019} as $\sigma_0=1.09191$, $b_s=15.36563$ and $p=3.0$. The temperature dependence is incorporated by
\begin{eqnarray}
h\bigg{(}\frac{T}{T_c(y_{c,p})}\bigg{)} &=& \bigg{[}1-\bigg{(}\frac{T}{T_c(y_{c,p})}\bigg{)}^2\bigg{]}^2
\hspace{0.5cm}\mbox{for}\hspace{0.2cm} T \le T_c(y_{c,p})\nonumber\\
&=& 0 \hspace{1.5cm}\mbox{for}\hspace{0.2cm} T > T_c(y_{c,p}).
  \end{eqnarray}
$T_c(y_{c,p})$ is the maximum temperature, for a given value of $y_{c,p}$ up to which nuclear liquid phase may co-exist with the nuclear vapour and it's expression in MeV unit is given by
\begin{eqnarray}
T_c(y_{c,p})=87.76\bigg{(}\frac{K_{sat}}{375}\bigg{)}^{\frac{1}{2}}\bigg{(}\frac{0.155}{\rho_0}\bigg{)}^{\frac{1}{3}}y_{c,p}(1-y_{c,p})
\end{eqnarray}
where $K_{sat}$, $n_0$ are expressed in MeV and fm$^{-3}$ respectively.
The Coulomb contribution in Helmholtz free energy is considered as same as before i.e.
\begin{eqnarray}
F^{Coul}&=&a^{*}_c\frac{Z^2}{A^{1/3}}
\label{Coulomb}
\end{eqnarray}
\begin{figure}[b]
\begin{center}
\includegraphics[width=\columnwidth]{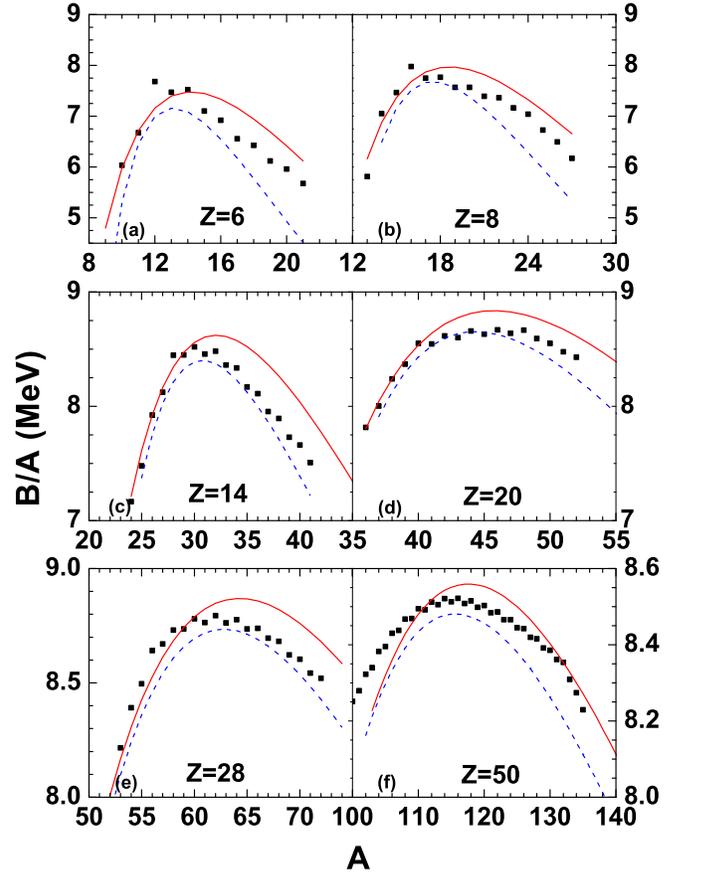}
\caption{Binding energy per nucleon of (a) Carbon (b) Oxygen (c) Silicon (d) Calcium (e) Nickel and (f) Tin isotopes obtained from nuclear liquid drop model (blue dotted lines) and realistic compressible liquid drop approach with Sly5 parameters (red solid lines). Experimental binding energies taken from AME2020 \cite{Wang} are shown black squares.}
\label{Binding}
\end{center}
\end{figure}
\section{Results}
\begin{figure}[!h]
\begin{center}
\includegraphics[width=\columnwidth]{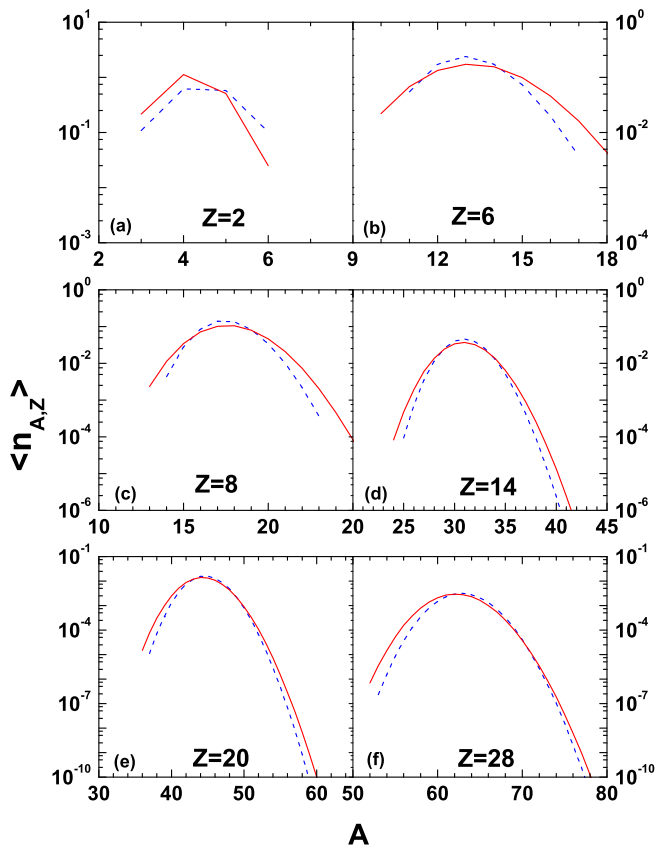}
\caption{Isotopic distributions for $Z$=2, 6, 8,14,20 and 28 from fragmentation of $Z_0$=75, $A_0$=168 system at $T$=5 $MeV$ studied from conventional CTM calculation (blue dashed lines) and CTM calculation with realistic Sly5 EoS (red solid lines).}
\label{Isotopic_A=168}
\end{center}
\end{figure}

\begin{figure}[t]
\begin{center}
\includegraphics[width=0.9\columnwidth]{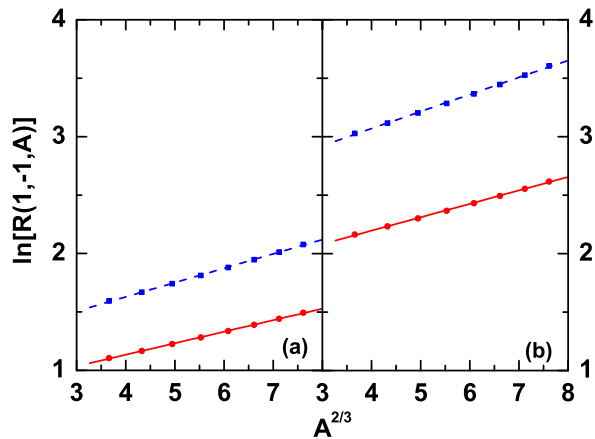}
\caption{Variation of isobaric yield ratio $ln R[1,-1,A]$ with $A^{2/3}$ for $Z_0$=75, $A_0$=168 (left panel) and $Z_0$=75, $A_0$=186 (right panel) from conventional CTM approach (blue dashed lines) and CTM calculation with realistic Sly5 EoS (red solid lines). Calculations are performed at temperature $T$=5 $MeV$ and freeze-out volume $V_f=6V_0$.}
\label{Isobaric_yield_ratio}
\end{center}
\end{figure}

\begin{figure}[!h]
\begin{center}
\includegraphics[width=0.9\columnwidth]{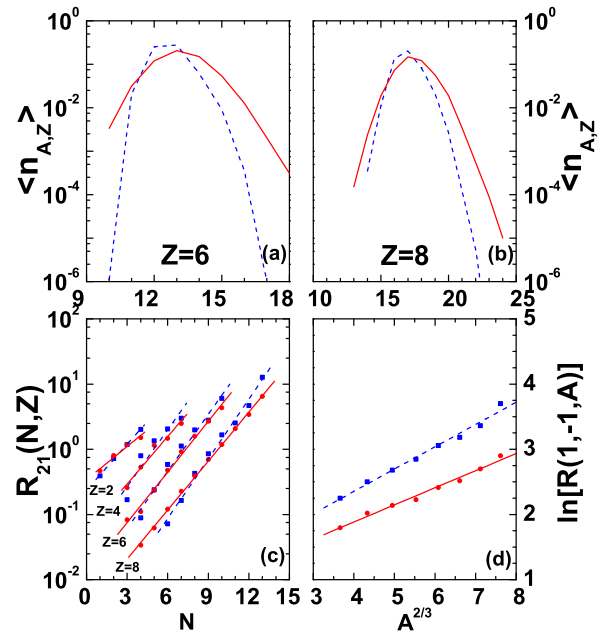}
\caption{Upper panels: Isotopic distributions for $Z$=6 (upper left panel) and  $Z$=8 (upper right panel) after secondary decay from $Z_0$=75, $A_0$=168 system. Lower left panel: Ratios($R_{21}$) of multiplicities of secondary fragments $(N,Z)$ where system 1 is $Z_0$=75, $A_0$=168 and system 2 is $Z_0$=75, $A_0$=186 (The lines drawn through the points are best fits of the calculated ratios.) Lower right panel:  Variation of isobaric yield ratio $ln R[1,-1,A]$ of secondary fragments with $A^{2/3}$ for $Z_0$=75, $A_0$=168. Blue dashed lines represents results from CTM calculation described in II-A followed by evaporation model  \cite{Mallik1} with binding from conventional liquid drop model and excitation from Fermi gas model where as red solid lines indicates results obtained from CTM calculation described in II-B followed by evaporation model with realistic binding and excitation.All calculations are performed at temperature $T$=5 MeV}
\label{Secondary_decay}
\end{center}
\end{figure}

Statistical model calculations are performed for two fragmenting systems having same proton number $Z_0$=75 but different mass numbers $A_0$=168 and 186 which are expected to be formed from the central collision of $^{112}$Sn+$^{112}$Sn and $^{124}$Sn+$^{124}$Sn reaction with $75\%$ pre-equilibrium emission \cite{Frankland,Xu}. The choice of fragmenting systems is based on well known experiments of $^{112}$Sn+$^{112}$Sn and $^{124}$Sn+$^{124}$Sn reactions at 50 MeV/nucleon performed by MSU group at NSCL \cite{Liu}. More precise calculation for identifying fragmenting source mass number, isospin asymmetry and excitation can be found in Ref. \cite{Tan, Mallik23}. As described in section 2, fragments with all possible proton number (i.e. $Z$=1 to 75) and neutron number (within neutron and proton dripline each $Z$) will be produced in multifragmentation reactions and their multiplicities are linked to the nuclear binding, therefore the binding energies of some selective cases like various isotopes of Carbon, Oxygen, Silicon,  Calcium, Nickel and Tin are shown in Fig. \ref{Binding}. Binding energy per nucleon of the above mentioned isotopes obtained from conventional liquid drop model and more realistic compressible liquid drop approach with Sly5 parameters are compared with their experimental values. From Fig. \ref{Binding} it can be concluded that the bindings are modified significantly for the neutron rich nuclei. Neutron rich nuclei are remarkably produced in multifragmentation reactions, hence the effect of this binding energy shift on the basic observables of multifragmentation reactions is described below.\\
\begin{figure}[!h]
\begin{center}
\includegraphics[width=0.8\columnwidth]{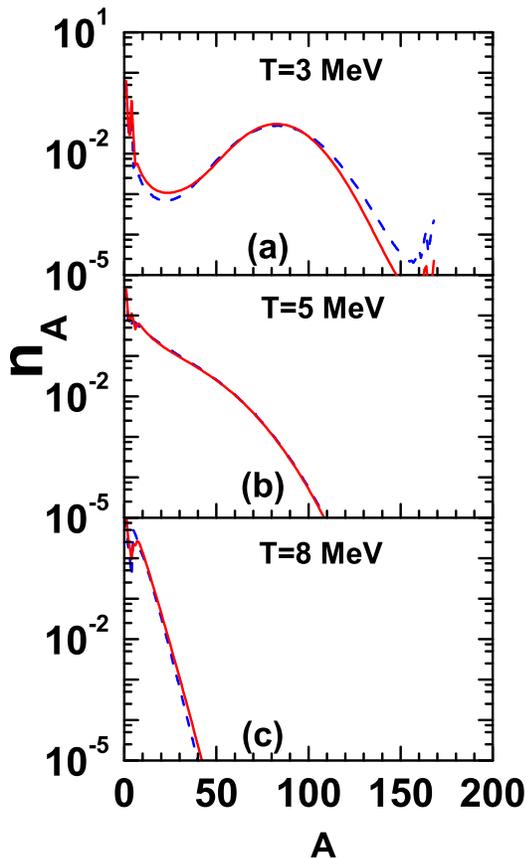}
\caption{Mass distribution from fragmentation of $Z_0$=75, $A_0$=168 system at temperature $T$=3 $MeV$(upper panel), 5 $MeV$(middle panel) and 8 $MeV$(lower panel) studied from conventional CTM calculation (blue dashed lines) and CTM calculation with realistic Sly5 EoS (red solid lines).}
\label{Mass_ditribution}
\end{center}
\end{figure}
\begin{figure}[!h]
\begin{center}
\includegraphics[width=0.8\columnwidth]{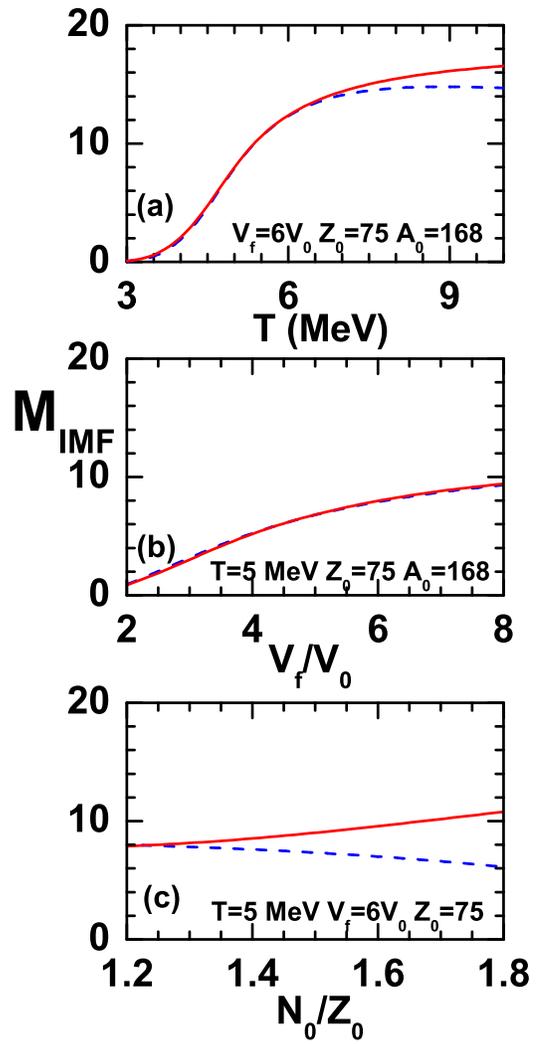}
\caption{Variation of multiplicity of intermediate mass fragments ($M_{IMF}$) with (a) temperature (upper panel) (b) freeze-out volume (middle panel) and isospin asymmetry of the fragmenting system (lower panel) from conventional CTM calculation (blue dashed lines) and CTM calculation with realistic Sly5 EoS (red solid lines). Calculations are performed for fragmenting system of atomic number $Z_0$=75 and mass number $A_0$=168.}
\label{Total_IMF_multiplicity}
\end{center}
\end{figure}
\indent
Fig. \ref{Mass_ditribution} represents the mass distribution obtained from CTM calculation with conventional liquid drop model and more realistic compressible liquid drop approach with Sly5 parameters at three different temperatures 3 MeV, 5 MeV and 8 MeV from disassembly of $Z_0$=75, $A_0$=168 at constant free-out volume 6$V_0$. The intermediate mass fragment (IMF) multiplicity ($M_{IMF}$) is also an important observable in nuclear multifragmentation process which is measured both experimentally and theoretically in many situations \cite{Peaslee,Ogilvie,Tsang,Ogul,Mallik3,Mallik16,Mallik19}. The IMF multiplicity ($M_{IMF}$) dependence with temperature (from disassembly of $Z_0$=75, $A_0$=186 at constant free-out volume 6$V_0$), freeze-out volume (for the same fragmenting system   $Z_0$=75, $A_0$=168 but at constant temperature 5 MeV) and neutron to proton ratio of the fragmenting system (from disassembly with constant $Z_0$=75 at constant temperature 5 MeV and free-out volume 6$V_0$) are presented in Fig. \ref{Total_IMF_multiplicity}. From Fig. \ref{Total_IMF_multiplicity}, it can be verified that, the effect of nuclear binding on intermediate mass fragment multiplicity is more significant at higher temperature and freeze-out volume i.e. the conditions at which the system breaks more. For a given temperature and freeze-out volume, the IMF multiplicity is strongly modified for the fragmentation from very isospin asymmetric systems due to significant production of neutron rich nuclei. Fig. \ref{Isotopic_A=168} represents  the isotopic distribution of some selective lower (Helium), intermediate (Carbon, Oxygen, Silicon and Calcium) and heavy (Nickel) mass fragments originated from the multifragmentation of $Z_0$=75, $A_0$=168 system at temperature 5 MeV and free-out volume 6$V_0$. As in IMF region, neutron rich isotopes are more bound in compressible liquid drop approach with Sly5 parameters compared to conventional liquid drop model, hence the production of neutron rich isotopes of Carbon, Oxygen, Silicon and Calcium is significantly enhanced.\\
\begin{figure}[b]
\begin{center}
\includegraphics[width=\columnwidth]{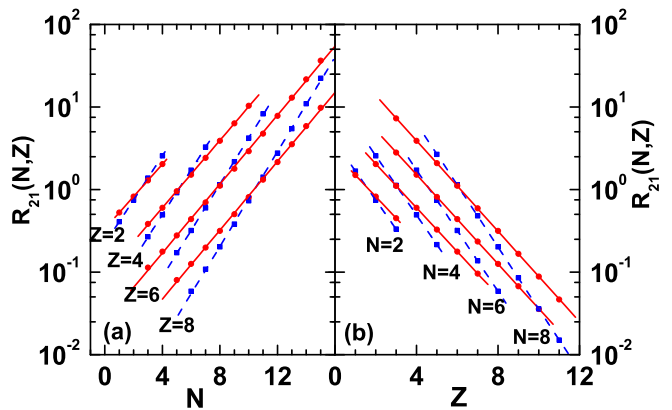}
\caption{Ratios($R_{21}$) of multiplicities of fragments of producing the nucleus
$(N,Z)$ where system 1 is $Z_0$=75, $A_0$=168 and system 2 is $Z_0$=75, $A_0$=186. Conventional CTM results (blue dashed lines) are compared with  CTM calculation with realistic Sly5 EoS (red solid lines). For each case reactions are simulated at constant temperature $T$=5 $MeV$ and freeze-out volume $V_f=6V_0$. The left panel shows the ratios as function of neutron number $N$ for fixed $Z$ values, while the right panel displays the ratios as function of proton number $Z$ for fixed neutron numbers.  The lines drawn through the points are best fits of the calculated ratios.}
\label{Isoscaling}
\end{center}
\end{figure}
\indent
One of the important aspect of intermediate energy heavy ion reactions study is to reduce the uncertainty in nuclear equation of state. Isoscaling \cite{Tan,Mallik23,Tsang1,Botvina1,Chaudhuri4,Colonna,Raduta,Mallik5,Mallik8} and isobaric yield ratio \cite{Mallik5,Mallik8,Huang,Tsang2,Ma1} methods are commonly used to search the precise nuclear equation of state. In this work, the effect of nuclear binding on isoscaling and isobaric yield ratio are investigated in the framework of CTM. It has been observed both experimentally and theoretically that the ratio of yields from two different reactions(having different isospin asymmetry), exhibit an exponential relationship as a function of the neutron($N$) and proton($Z$) number and this is termed as 'isoscaling'. Two fragmentation reactions "$1$" and "$2$" at a given energy are being considered whose fragmenting systems have different mass $A_{01}$ and $A_{02}$ ($A_{02}>A_{01}$) but same charge $Z_1=Z_2=Z_0$.
\begin{eqnarray}
R_{21}&=& \langle {n_2}_{N,Z}\rangle/\langle {n_{1}}_{N,Z}\rangle\nonumber\\
&=& C\exp(\alpha N+\beta Z)
\label{isoscaling_eq}
\end{eqnarray}
Where $\alpha$ and $\beta$ are the isoscaling parameters and $C$ is a normalization factor. Fig. \ref{Isoscaling} represents, the isoscaling ratios obtained from CTM calculation at constant temperature $T=5$ MeV and freeze-out volume $V_f=6V_0$ for two fragmenting sources with identical atomic number $Z_0=75$ but different mass number $A_{01}=168$ and $A_{02}=186$. The isoscaling ratio $R_{21}$ is plotted as function of the neutron number ($N$) for $Z=2$, $4$, $6$ and $8$  in the left panel whereas the right panel displays the ratio as function of the proton number ($Z$) for $N=2$, $4$, $6$ and $8$. The lines are the best fits of the calculated $R_{21}$ ratios to Eq. \ref{isoscaling_eq}. From Fig. \ref{Isoscaling}, it can be concluded that CTM calculation with more realistic semi-microscopic binding and excitation also shows isoscaling behavior but with the inclusion of realistic semi-microscopic binding, the magnitude of the isoscaling parameters $\alpha$ and $\beta$ decreased.\\
\indent
The isobaric ratio of yields \cite{Huang} of two different types of fragments having same mass number $A$ but different isospin asymmetry $I=N-Z$ and $I^{'}=N^{'}-Z^{'}$ originating from the fragmenting system is given by,
\begin{equation}
R[I^{'},I,A]=\langle n_{I,A}\rangle/\langle n_{I^{'},A}\rangle
\label{Isobaric_yield_ratio_eq}
\end{equation}
The quantity $R[I^{'},I,A]$ shows linear behavior with $A^{2/3}$ for $I$=1 and $I'$=-1 from conventional liquid drop approach \cite{Mallik8}. Fig. \ref{Isobaric_yield_ratio} confirms that, for semi-microscopic realistic binding this linear behavior also holds but for a given fragment mass number, production of neutron rich isotopes is more with realistic binding compare to conventional liquid drop approach, hence the isobaric yield ratio value is less. This reduction is more for fragmentation from more isospin asymmetric system ($Z_{0}=75$  and $A_{0}=186$) compare to other one ($Z_{0}=75$  and $A_{0}=168$).\\
\indent
The excited fragments produced in the multifragmentation stage decay to their stable ground states. They can $\gamma$-decay to shed energy but may also decay by light particle emission to lower mass nuclei. Hence an evaporation model with the same realistic semi-microscopic binding and excitation is developed. Emissions of $n,p,d,t,^3$He and $^4$He particles are considered. Particle-decay widths are obtained using Weisskopf's evaporation theory \cite{Weisskopf}. To study the cluster functional effect on cold fragments for isospin sensitive observables like isotopic distribution, isoscaling and isobaric yield ratio, two separate cases are considered-(a) CTM calculation described in II-A followed by evaporation model  \cite{Mallik1,Mallik_Thesis} with binding from conventional liquid drop model and excitation from Fermi gas model and (b) CTM calculation described in II-B followed by evaporation model with realistic binding and excitation. The results are presented in Fig. \ref{Secondary_decay},  which confirms relative enhancement of multiplicities of very neutron rich isotopes and reduction of isoscaling parameters as well as isobaric yield ratio due to introduction of new cluster functional exist strongly even after secondary decay.

\section{Summary and future outlook}
The canonical thermodynamical model of nuclear multifragmentation is upgraded with more realistic semi-microscopic binding and excitation. The bulk part of the binding is determined from newly proposed meta-modelling of the equation of state with Sly5 parameters. The effect of cluster functional has been examined for basic obseravables of nuclear multifragmentation like intermediate mass fragment multiplicity and mass distribution as well as isospin sensitive observables like isotopic distribution, isoscaling and isobaric yield ratio at different thermodynamic conditions of temperature and freeze-out volume and isospin asymmetry of fragmenting system which can be accessed in laboratory experiments. Semi-microscopic realistic binding significantly modifies intermediate mass fragment production (from isospin asymmetric fragmenting systems), multiplicity of neutron rich isotopes as well as isoscaling, isobaric yield ratio parameters. These modifications due to inclusion realistic cluster functional also present after the secondary decay of the excited fragments.\\
\indent
In a future work it will be interesting to study the effect of different nuclear EoS of fragmentation observables in the framework of this upgraded statistical model. Hybrid model calculations are quite successful for explaining projectile fragmentation reactions in the limiting fragmentation region, but one of the major constraint of presently available hybrid models is that the dynamical stage is performed with microscopic interaction where as fragmentation is treated with statistical models where conventional liquid drop binding and level density are used, hence substantial error may arise due to this inconsistency during the coupling of the transport model output result and further statistical model calculation, this newly proposed CTM model will bypass such difficulty and can be coupled more efficiently with dynamical model with identical interaction in both stages. It will be very interesting to pursue that in a future work.
\section{Acknowledgement}
The author gratefully acknowledge Francesca Gulminelli of LPC Caen, Gargi Chaudhuri of VECC and Subal Das Gupta of McGill University for valuable discussions and suggestions.

\end{document}